# Design of materials properties and device performance in memristive systems

M. Lübben[1,2], F. Cüppers[1,2], J. Mohr[1,2], M. von Witzleben[1,2], U. Breuer[3], R. Waser[1,2,4], C. Neumann[5*], I. Valov[2,4*]

[1]Institut für Werkstoffe der Elektrotechnik II, RWTH Aachen University, Sommerfeldstraße 24, 52074 Aachen, Germany

[2]JARA – Fundamentals for Future Information Technology, 52425 Jülich, Germany

[3]Central Institute for Engineering, Electronics and Analytics, Forschungszentrum Jülich GmbH, 52425, Jülich, Germany

[4]Peter-Grünberg-Institut (PGI 7), Forschungszentrum Jülich, Wilhelm-Johnen-Straße, 52425 Jülich, Germany

[5]Heraeus Deutschland GmbH & Co. KG

## Abstract

Future development of the modern nanoelectronics and its flagships internet of things and artificial intelligence as well as many related applications is largely associated with memristive elements. This technology offers a broad spectrum of functionalities, however, it follows predominantly a phenomenological approach and crucial challenge/limit for further development remains variability and lack of fundamental materials' design strategy.

Here we demonstrate the vital importance of materials' purity for determining memristors' functionalities, showing that part per million foreign elements significantly change the performance. By appropriate choice of chemistry and amount of doping material we can selectively enhance desired operation mode. We highlight how dopant dependent structure and charge/potential distribution in the space charge layers and the cell capacitance determine the device kinetics and functions. We evidence for first time experimentally the relation between materials properties and switching/neuromorphic performance, thus providing rules and directions for a rational design of memristive devices.

## Introduction

Memristive cells and devices are essential building units for future nanoelectronic architectures targeting alternative data processing paradigms such as cognitive/neuromorphic computing and alternative logic operations, all being milestones in assembles of global networks such as Internet of Things (IoT) and creation of artificial intelligence (AI). Memristive cells are energy efficient, scalable to almost atomic level, allowing read/write/erase operations within less than nanoseconds. These devices demonstrate multiple functionalities, such as non-volatile memory and selector functions, ability to transit from digital to analog data storage and processing, sensor activities etc.[1, 2] They are also extremely stable against high energy particles and electromagnetic waves, thus preventing soft errors and can operate in a unprecedently broad temperature range from 4 K[3, 4] to 1000 K[5] making their application in space technologies and in general application under harsh conditions unavoidable.

In the recent years, the application of memristive devices evidenced remarkable progress in areas of downscaling, pattern/face recognition, integration in chips, arrays and complex architectures[1, 6-11]. However, in most cases the materials combination within the device structure are selected on empirical rules, depending on the particular operation type and/or the specific application. Furthermore, in many cases same materials are used for different applications without identifying the reasons for the different performances. Thus, an essential fundamental aspect has still not been addressed – the relation between the design of intrinsic material properties, and device performance and functionalities.

The era of the silicon-based industry has begun after realizing the importance of extreme purity required for Si/Ge components. In the 60's Siemens researchers succeeded in purification of Si by zone melting [12] to reduce residual impurities. Doping strategies have been developed to enhance one or other electronic property in these high purity semiconductor materials. The issue of materials' purity has not been considered yet for memristive devices.

This study highlights the relation between material purity, chemistry and concentrations of intrinsic/extrinsic doping, and device performance and functionalities. We evaluate and demonstrate how material properties and space charge layer structure change upon doping $SiO_2$ of highest purity with various elements, and accordingly how the device properties are modulated. The levels of doping we use are below the impurity levels of materials typically used for memristive devices. Depending on the type and amount of the doping element we can enhance or suppress particular characteristics such as fast/slow switching, short/long retention time, short term plasticity (STP) or long term plasticity (LTP), potentiation etc.

Our results apply to all type of redox-based memristive devices and materials and provide general design rules for selective device improvement.

**Results & Discussion**

**Influence of doping and moisture on dielectric/defect properties**

The physicochemical properties of amorphous oxide materials used for memristors are typically discussed in the context of point defects in crystalline solids such as oxygen vacancies, interstitial ions etc. Despite this approach is widely spread in memristive community and appears in many cases useful, there are several important arguments that limit this approximation. i) Definition of point defects is strictly valid only for crystalline solids with a periodic lattice order. ii) Point defect dynamics can be difficultly compared at low (e.g. room) and high temperatures. iii) At lower temperatures point defects are usually associated within the oxide matrix, forming dipoles, which has not yet been considered.

We propose an effective alternative treatment, based on determination of defect-chemical concentration and properties by measuring the dielectric characteristics of the nanoscale films and thus, estimating the impurities/dopant concentrations.

We selected for model study $SiO_2$ as switching material in electrochemical metallization (ECM) device of the type $Cu/SiO_2/Pt$. This system is extensively studied and underlying electrochemical processes and switching kinetics are well documented and understood[13-19]. We used as a reference material pure $SiO_2$ (purity degree: 8N+). Copper (to provide mobile ions), aluminum and gallium in different combinations were added to distinguish and highlight the effects of the dopants. Unlike other approaches using thermal diffusion[20, 21] or photodiffusion[22], we used sputter targets with defined compositions, allowing us to have highly uniform concentration profiles throughout the switching layer (Supplementary Note 1).

Typical defects in $SiO_2$ are protons and hydroxyl groups, that can be weakly bonded (physisorption), tightly (chemisorption) or directly incorporated in the Si-O-network[23], oxygen-deficiency centers, dangling bonds, interstitial oxygen and several more options (e.g. see [24]).

Dopants in the SiO$_2$ matrix (starting from ppm range) change its dielectric properties. Because of the different electron affinities of the foreign elements and/or defects, stronger or weaker interatomic electrostatic interactions are induced, leading to formation of dipoles, changing the permittivity and capacitance of the film.

In Figure 1, pure and differently doped SiO$_2$ films and their permittivity are plotted as a function of non-volatile (Cu, Al, Ga) and volatile (moisture uptake in SiO$_2$) doping concentrations.

The permittivity of both pure and doped SiO$_2$ films increases with increasing humidity levels following a linear relation (Supplementary Note 2), where for very high $p_{H2O}$ the uptake of moisture begins to saturate.

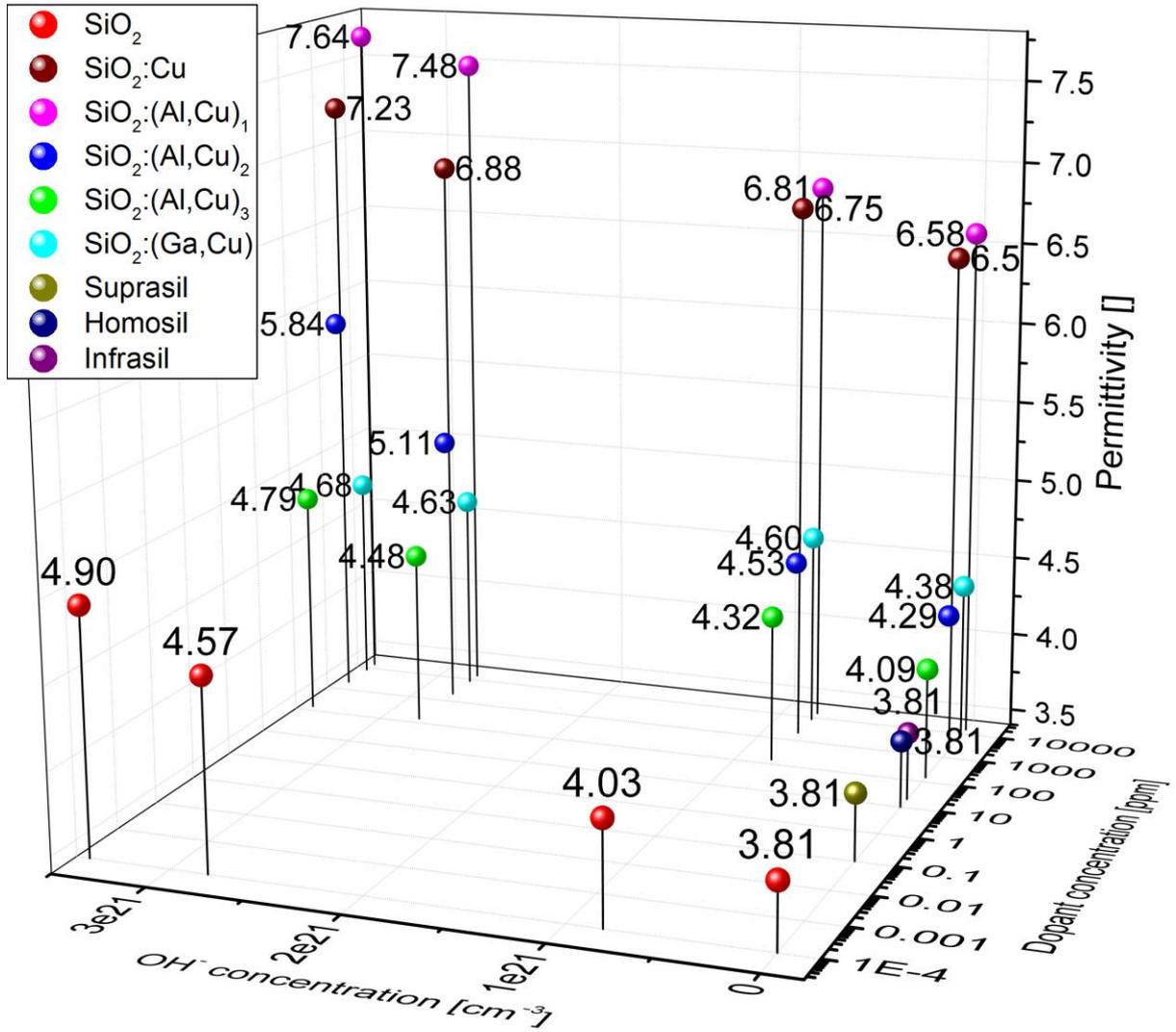

Figure 1: Permittivity values for pure and doped SiO$_2$ as a function of OH$^-$ concentration (water content) and dopant concentration. Highly pure SiO$_2$ resembles water free undoped Suprasil W. The permittivity values for the sputtered samples were measured in vacuum, at ~35 % relative humidity and >90 % relative humidity. We calculated the residual OH$^-$ concentration in SiO$_2$ films according to [25]. This calculation procedure was also used for doped SiO$_2$. For all materials, the permittivity is increasing with increasing OH$^-$ (humidity) concentration, following linear relation $\varepsilon_s = 3.8073 + 2.72 * 10^{-22} \frac{cm^3}{ion} * N$ [25], where $N$ is the concentration of OH$^-$ defects per cm$^3$. The increase in Al content leads to a much higher polarizability of the material, whereas Ga doping has lower impact on $\varepsilon$. The concentrations and permittivities of the glasses Suprasil, Homosil and Infrasil are extracted from [26], [27] and [25]. The method for determining dopant concertation by dielectric measurements has a sensitivity even in 1 ppm range (Supplementary Note 2).

Extrinsic dopants e.g. Al or Ga additionally increase the dielectric constant. The increase of the permittivity origins in the formation of dipoles between Al-ion (relative charge in Kroeger-Vink notation $Al'_{Si}$) that can be considered acceptor and Cu-ion ($Cu^{\bullet}_I$ or $Cu^{\bullet\bullet}_I$) being a donor as schematically shown in Figure 2a. This value however, is not solely determined by the absolute concentration and charge of the particular species, but accounts also for the chemical nature of the dopants, for the interactions with surrounding matrix and other dopants.

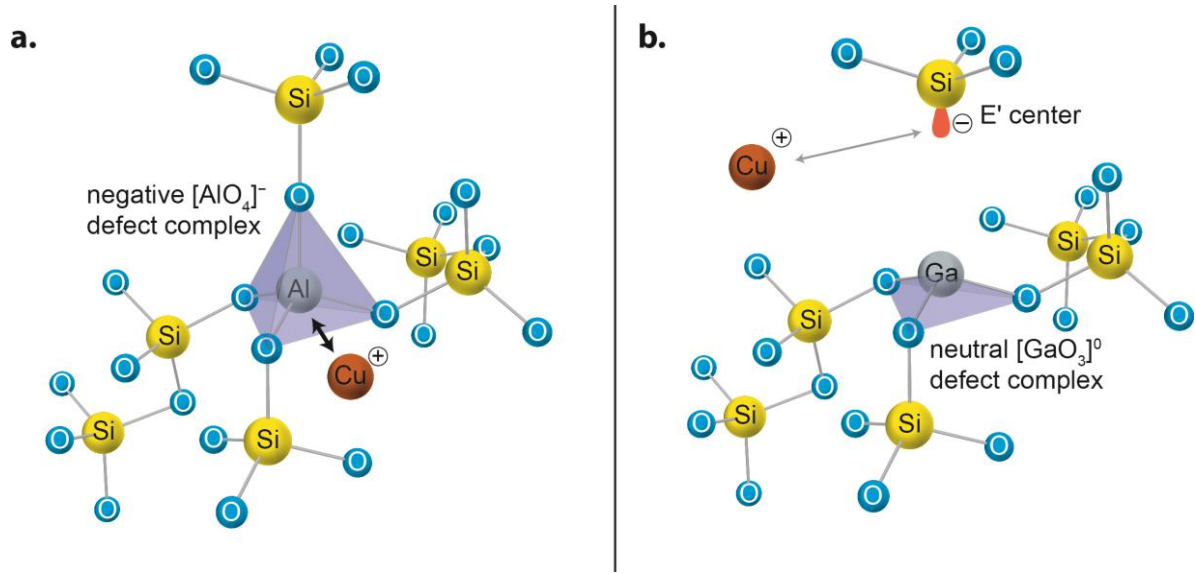

Figure 2: Schema of the atomistic configuration in $SiO_2$ amorphous films around a substitutional atom. a) Al ion replaces a Si ion. Because of the electron configuration of the Al, the local charge surplus can bind $Cu^+/Cu^{2+}$ ions in the vicinity of the Al atom. b) Ga substitution. Ga is forming on $[GaO_3]^0$ complex and Cu-ions interacts only with the E´ center of the $[SiO_3]^-$. This bond is weaker compared to that between Cu and $[AlO_4]^-$.

For example replacing Al in $SiO_2$:(Al,Cu)$_1$ by same amount of Ga results in a lower permittivity of $\varepsilon_{rSiO2:(Ga,Cu)} = 4.60$ (compared to 6.81), because Ga forms a neutral defect complex. A direct correlation of the absorption band at 215 nm wavelength observed in Ga-doped $SiO_2$ with increasing doping concentration suggests a defect complex with an associated negatively charged E' center[28]. Therefore, Ga only indirectly forms a different negatively charged defect center and thus creates different defects in $SiO_2$ than aluminum with its $AlO_4^-$ configuration[24] as shown in Figure 2b with its strongly bonded dipoles with Cu. Still, Ga has same oxidation state as Al and also leads to the same amount of negatively charged defect complexes - but with strongly differing dipolar interaction to dissolved Cu ions.

Dopants are also interacting with moisture. For example, whereas the slopes of permittivity vs. $p_{H2O}$ dependence for (Al,Cu)-doped samples are relatively steep (Supplementary Figure S5), the slope for the (Ga,Cu)-doped sample is much flatter, indicating a weaker interaction of the matrix with moisture, due to chemical saturation effects induced by Ga-doping. Therefore, by properly selecting the doping element we can design a material, respectively a memristive cell, with stronger or weaker interactions/bonding between mobile and immobile defects (resulting in different transport properties, switching kinetics and filament stability) and robustness against humidity and/or other external influences[29].

**Nanoscale field enhanced redox reactions and transport**

Thickness and structure of the electrochemical double layers (EDL) are of special importance for the switching kinetics, filament stability and dynamics in memristive devices. In nanoscale systems, double layers may overlap at conditions, fulfilled in most cases by memristors. Such an overlap can electrostatically hinder or enhance the reaction rate and distribution/movement of charged species, thus modifying the device kinetics and stability of ON, OFF and intermediate states.

We calculated in Figure 3 the thickness of EDL in pure and doped $SiO_2$ and the resulting electric potential/field distribution for a symmetric $Pt/SiO_2/Pt$ and an asymmetric $Cu/SiO_2/Pt$ stack, respectively. For undoped $SiO_2$ the Debye length is calculated $\lambda_{D,undoped} = 69$ µm and the electric potential cannot relax over the oxide thickness (10 nm). The double layers overlap and the electric field drops linearly over the entire film thickness. For higher dopant/impurity concentrations $\lambda_D$ becomes smaller and the potential/field distribution takes its classical form.

The electrical potential gradient in memristive cells can have intrinsic or extrinsic nature and determines the distribution of mobile charges within the space charge layer and the reaction/switching kinetics. For example, $Cu/SiO_2/Pt$ devices have an internal electromotive force $\Delta\varphi_{int}$ (nanobattery effect) that may reach over 400 mV [13]. This $\Delta\varphi_{int}$ leads to a charge separation and formation of charge(ion) clouds of opposite sign at electrode/electrolyte interfaces and the formation of a depleted intermediate region (more insulating) as shown in Figure 3c. In case of a single mobile type charge, one of the interfaces will be enriched and the other, respectively depleted of charge. Similar considerations apply to valence change memories (Supplementary Note 3) or other devices, where defects form associates at room temperature[30, 31]. The internal rearrangements make the cell similar to electrochemical capacitor, where the capacitance value depends on $\Delta\varphi_{int}$, the dielectric constant of the pure electrolyte material, its thickness and the initial concentration of mobile charges $N_0$. The resulting concentration $N$ profiles as shown in Figure 3c follow exponential dependence:

$$N = N_0 na \exp\left(\frac{ze}{k_B T}\Delta\varphi_{int}\right) \qquad (1)$$

with $a$ being the jump distance corresponding to the distance between two monolayer planes of the solid electrolyte and $n$ is the total number of these planes ($n \cdot a$ is giving the thickness of the oxide film $d$); $z$ is the charge number; $e$ is the electron charge; $k_B$ and $T$ are the Boltzmann constant and temperature, respectively.

These specifics of nanoscale cells have decisive influence on the switching kinetics. Depending on the sign of $\Delta\varphi_{int}$, it will be either added (accelerating effect) or subtracted (suppressing effect) to/from applied bias during device operation, thus significantly changing the reaction rate and switching time. The effect is especially pronounced for overlapping double layers where both $\Delta\varphi_{int}$ and applied voltage $\Delta\varphi_{appl}$ lead to field dependent reaction kinetics and transport:

$$i = N\nu z e \exp\left(-\frac{\Delta G_a^{\neq} - \alpha a z e}{k_B T} \frac{\Delta\varphi_{appl} \pm \Delta\varphi_{int}}{d}\right) = i_0 \exp\left(-\frac{\alpha a z e}{k_B T} \frac{\Delta\varphi_{appl} \pm \Delta\varphi_{int}}{d}\right) \qquad (2)$$

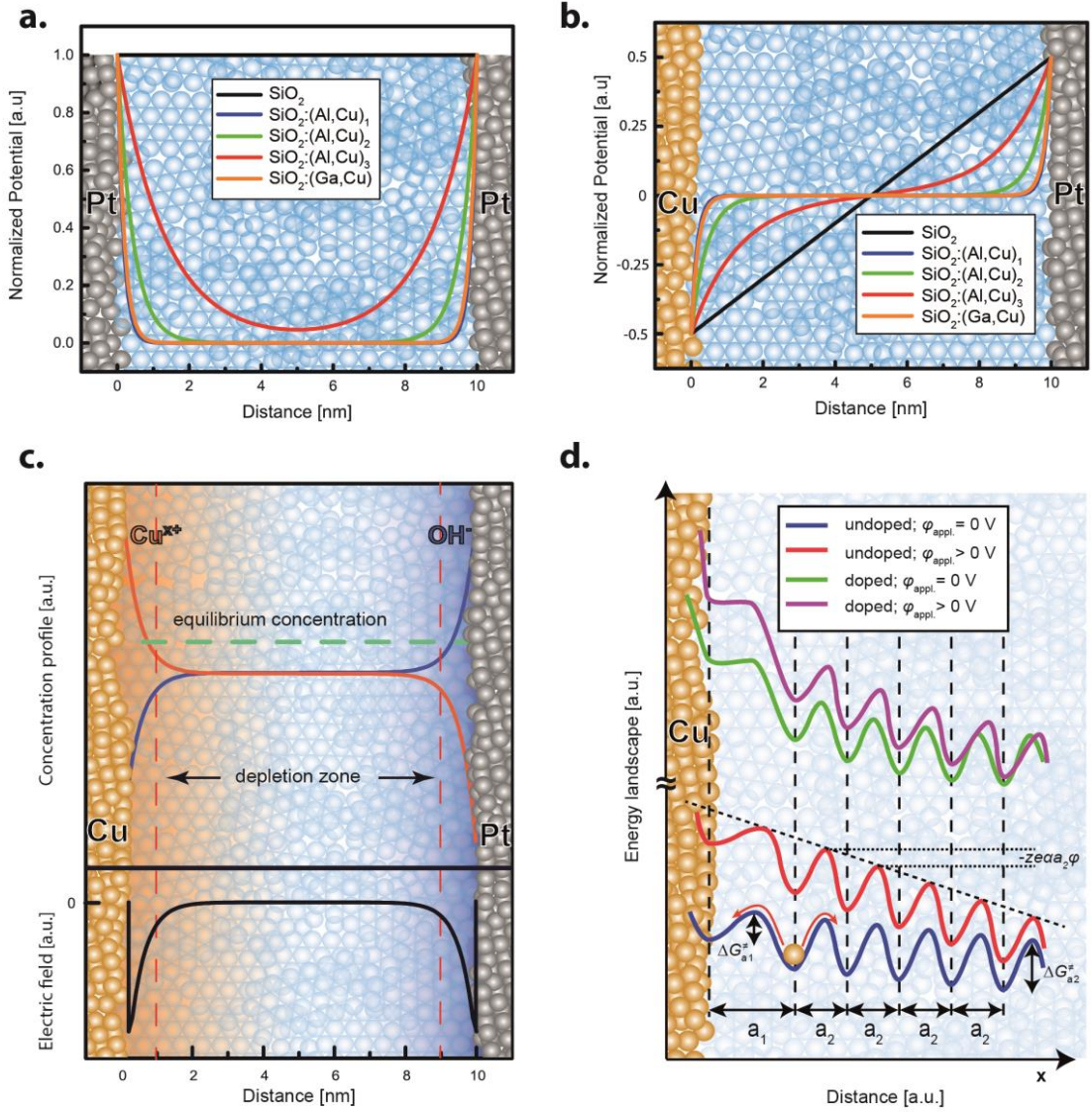

Figure 3: Potential and energy distribution in nanoelectrochemical memristors. a) Symmetric cell with Pt electrodes b) Asymmetric cells with Cu and Pt as electrodes. Details on the calculation of the Debye length and the corresponding values are given in Supplementary Note 4. c) Charge separation and formation of ion enriched and depleted zones in asymmetric memristive devices with two mobile charge species and the corresponding electric field distribution (or SiO$_2$:(Al,Cu)$_2$. In the particular case the electric field drops only within the small zones of higher charge concentration. In case of pure material the field will drop across the entire film thickness (see b)). d) Schematic sketch of the energy landscape for pure and doped

samples with and without applied external voltage. The difference between doped and pure $SiO_2$ is presented by the change in the principle shape of the energy potential curve. Charge separation can additionally occur under applied external bias in cases that jump distances $a_1$ and $a_2$ and/or activation energies $\Delta G^{\neq}_{a1} / \Delta G^{\neq}_{a2}$ for redox reactions and transport respectively, differ.

with $i$ being the current density, $\Delta G^{\neq}_a$ is the activation energy, $\alpha$ is the transfer coefficient and $i_0$ is the exchange current density.

Equation 2 can be formally applied to both charge transfer and transport limited kinetics. The difference will be only in the particular values for the activation energy $\Delta G^{\neq}_a$, the jump distance $a$ and the resulting charge transfer coefficient $\alpha$. These factors can be alone a reason for formation of regions with locally uncompensated charges (Supplementary Note 5).

Adding immobile dopants to $SiO_2$, i.e. Al or Ga, induces electrostatic interactions between the dopant(s) and mobile ions, leading to less pronounced charge separation, changes the potential/field distribution and all kinetic parameters, but especially the activation energy for transport, increased by the energy for breaking the bond to the doping element (Fig. 3d).

**Switching Kinetics**

The switching kinetics and the type of device functionality can be pre-determined and/or adjusted by the type/chemistry and levels of doping in the $SiO_2$ matrix.

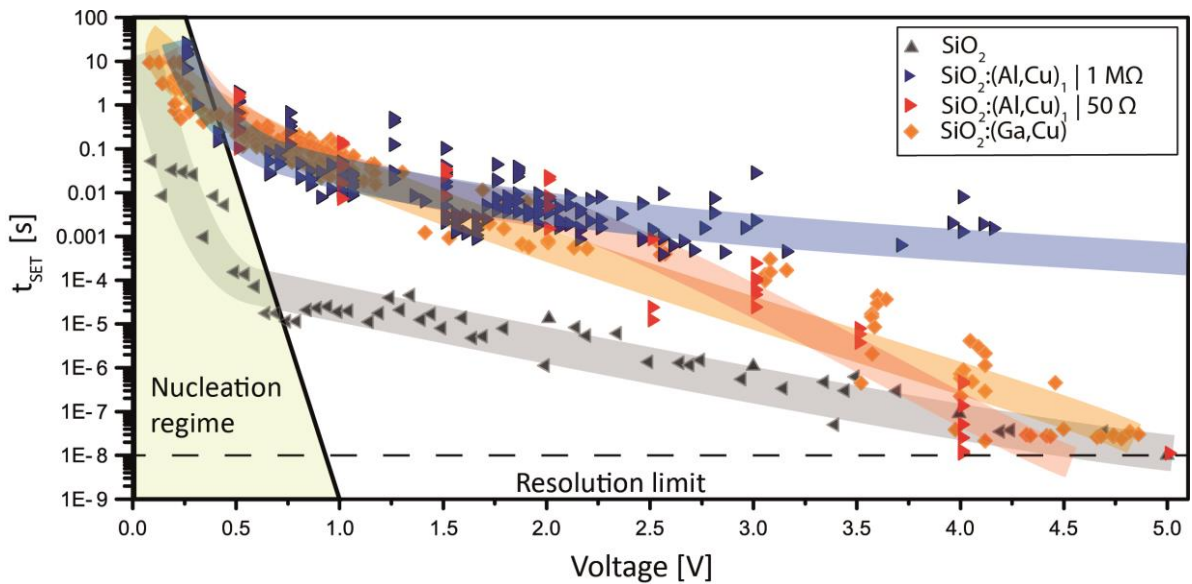

Figure 4: SET kinetics for $SiO_2$-based devices. In the low voltage regime the SET kinetics are limited by the nucleation time and the slope of the kinetics is steep. For higher voltages, the time determining mechanism is strongly dependent on the doping and its influence on the switching properties. The undoped device (grey) show the typical trend for an ECM switching kinetic with a pronounced steep nucleation regime followed by a flatter slope representing a limitation by the ion movement and/or redox processes[32]. The dependence of the SET Kinetics on the moisture levels was demonstrated in [33]. For the highly doped $SiO_2$:$(Al,Cu)_1$ devices with the highest permittivity the kinetics are much slower and additionally limited by the RC time given by the high capacity and the shunt resistance in the setup. With a 1 MΩ resistance the switching speed is limited to times of around 1 ms at 5 V (blue), see also Supplementary Note 6. When measuring with 50 Ω impedance, the kinetics are similar up to 2 V. Above 2 V the kinetics with the low impedance are becoming faster than with high impedance. The SET kinetic shows a plateau at voltages lower than 2 V. This we account to the stronger interaction between the doped $SiO_2$ (especially the $AlO_4^-$-clusters) and the $Cu^{2+}$-ions (formation of defect associations, see also Figure 2) involved in the switching

Undoped $SiO_2$ shows the fastest SET times being orders of magnitude lower compared to doped samples. In Figure 4, pronounced nucleation regime and overlapping charge transfer and diffusion limited kinetics can be observed. By reducing the $SiO_2$ film thickness to 3 nm, we achieved $t_{SET}$ = 1.4 ns being the fastest switching time for ECM devices reported up to now (Supplementary Note 7). The reason for this fast kinetics is the purity of the material. As it can be seen from Figure 3, for pure $SiO_2$, the space charge layers overlap and the kinetics is exponentially enhanced by the electric field within the entire film thickness. Due to the low $N_0$ there is no strongly pronounced charge separation effect and the electric field is almost homogeneously distributed within the SiO$_2$. No electrostatic interaction with doping elements is present. The external voltage is in this case superimposed to the in-built voltage, resulting in extremely fast switching time.

Highly doped samples show much slower kinetics. The nucleation limited regime for SiO$_2$:(Al,Cu)$_1$ devices is subtle, followed by a plateau in the voltage range 1 V – 2 V. We propose, this plateau is caused by the additional energy that is needed to break the bond between copper and AlO$_4^-$ clusters. Only after this activation energy is overcome Cu-ions can further respond to the voltage signal and move/switch faster. This effect is highlighting the importance of the interactions between incorporated dopants for the device kinetics.

The region above 2 V is characterized by a mixed charge transfer and diffusion limitation, logarithmically dependent on time. Finally for voltages above ~ 4 V field effects are too strong and SET time for doped and undoped SiO$_2$ becomes equal.

Samples with lower Al and Cu concentrations i.e. SiO$_2$:(Al,Cu)$_2$ and SiO$_2$:(Al,Cu)$_3$ show SET kinetics transiting from low to high doping regime (Supplementary Note 8). Evidently, even small amounts of dopants and/or impurities have significant influence on the switching.

The results for SiO$_2$:(Ga,Cu) are following the conclusions derived for the Al dopant, but due to weaker Ga-to-Cu electrostatic interactions, SiO$_2$:(Ga,Cu) devices show lower permittivity compared to SiO$_2$:(Al,Cu)$_1$ leading to smaller capacities, but more importantly the copper ions can be easily separated from Ga-ions leading to a weaker electrochemical capacitance and therefore faster switching.

**Neuromorphic functionalities**

Memristive devices are used to emulate neuromorphic functions, building artificial neural networks for "deep learning", pattern recognition, signal processing and brain-inspired spiking based computation[9, 34-39]. Key feature is the ability to imitate synaptic plasticity, i.e. to learn/forget depending on the amplitude and frequency of input signal. In biological systems this change (synaptic weight) relies on the modulation of cation (e.g. $Ca^{2+}$, $Na^+$, $K^+$) conduction channels by electrical stimuli[40], being categorized in two time scales: Short-term plasticity (STP) and long-term plasticity (LTP)[41]. In STP, the change is volatile and the synaptic weight relaxes to its ground state. The LTP is non-volatile and commonly represents the long-term memory function of a synapse. The transition between short- and long-term plasticity is fluent, depending on pulse amplitude, length of the pulses, length of pulse delays and the history of synapse/device itself. A synapse/device should be able to be potentiated (ON state) with positive pulses and depressed with negative pulses (OFF state).

Doping of the oxide is the most efficient way to modulate the potentiation and depression processes and control the pulse/delay window by tuning permittivity and electrochemical capacitance.

In Figure 5 a pulse train (0.8 V amplitude, 100 µs pulses, 100 µs delay) and the corresponding response of $SiO_2$:$(Al,Cu)_1$ device is shown, divided in three regions. In the first region, the cell requires a certain number of potentiation pulses, respectively time to change its electrical conductivity. The incubation time of 48 ms, is determined solely by charging and discharging the capacity, controlled by the type and amount of doping elements. Time and driving force are still not sufficient for Cu-ions to migrate through the oxide film and form a filament.

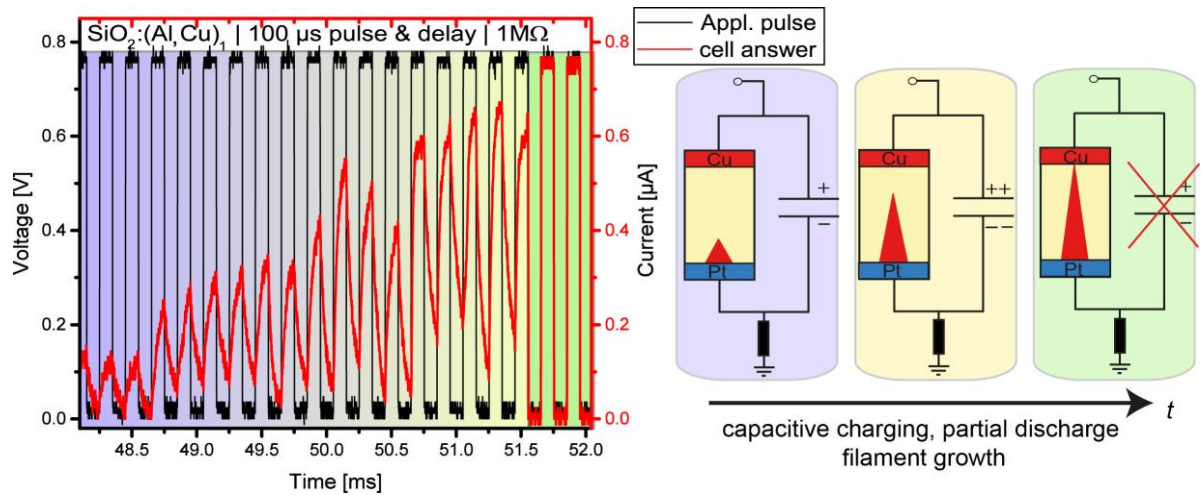

Figure 5 Potentiation of $SiO_2$:(Al, Cu)$_1$ memristive device. Red curve is the current response to the input voltage signal (black curve). As long as the switching filament has not short circuited the device, the capacity is loaded and discharges in every pulse and following delay time (first region, marked in blue). In the second region (yellow) the filament continues to grow. As long as charge is stored in the capacity, a driving force for further filament growth is present. In the event of a fully formed filament (third region, marked in green), the capacity in shorted and no further charging/discharging can be distinguished. The device's capacity plays a significant role in the potentiation process for neuromorphic applications influencing short term plasticity component which can learn and forget in short time ranges, but also the transition to long term memory. The electrical circuit during potentiation is present at the right side in the figure.

In the second region Cu-ion reduction starts at the Pt electrode and the conductivity of the device consequently increases due to filament growth until short circuit condition is reached. The final stage of the filament growth is clearly evidenced in the $V$ vs $t$ dependence in Figure 5 at 51.5 ms.

The dopant controlled capacity plays an essential role for the learning mechanism. On one hand side it slows down the kinetics of filament formation, but on the other hand side it is ensuring higher stability of the ON/Intermediate states, allowing to play between these two parameters which appears extremely important for neuromorphic applications. Moreover, during delay periods the capacity (including electrochemical capacity) acts as an additional voltage source and enhance the reduction rate of $Cu^{2+}$ ions and can initiate formation and/or hinder the dissolution of partially formed filament (nanobattery effect[42]).

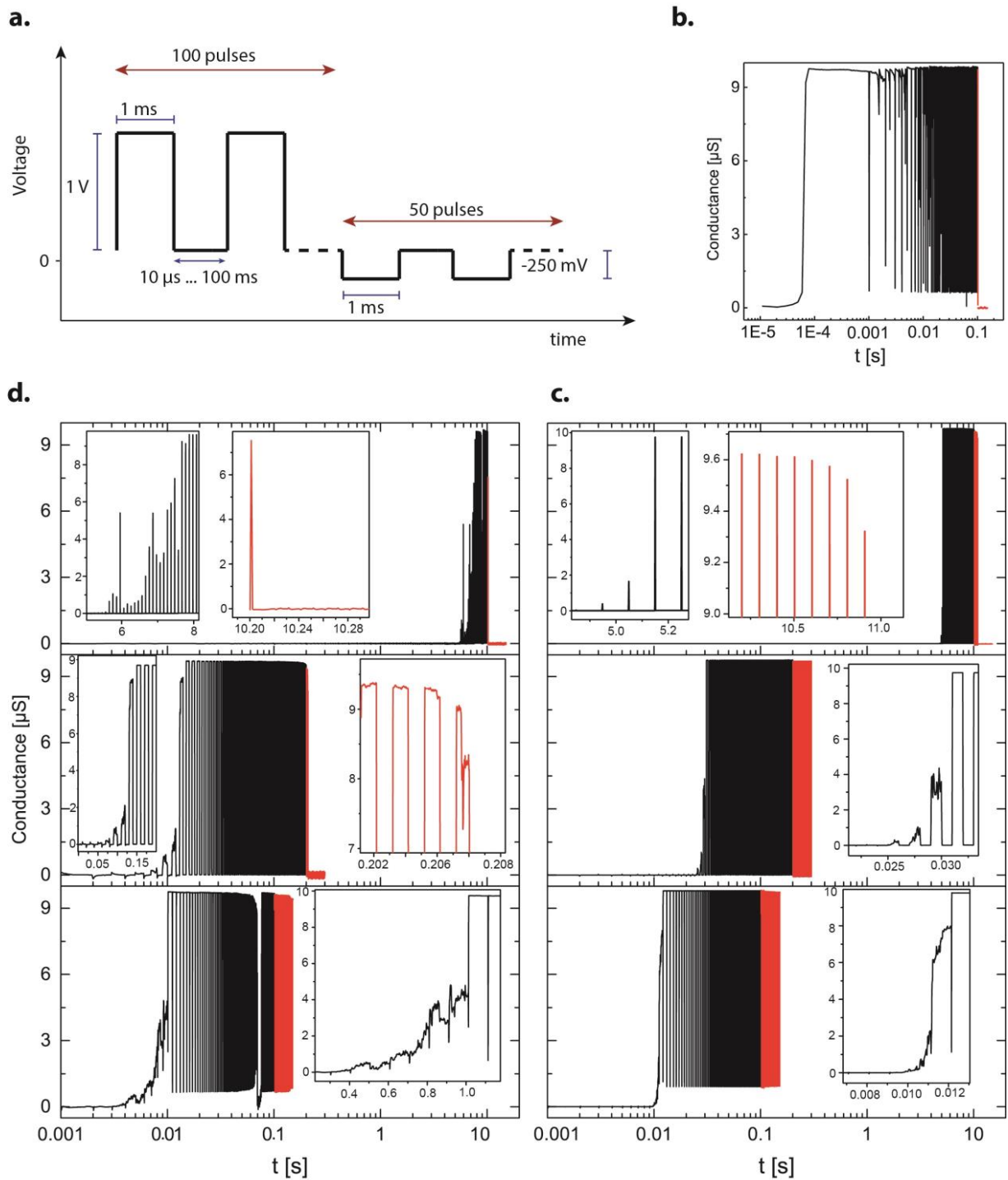

Figure 6: Potentiation and Depression response of SiO$_2$-based memristive devices. a) The pulse scheme applied to the cells. b) undoped SiO$_2$ c) highly (Al,Cu)-doped SiO$_2$ and d) (Ga,Cu)-doped SiO$_2$. The delay between the pulses was varied between 10 μs and 100 ms to simulate different presynaptic firing rates. A resistor of 100 kΩ connected in series limits the current to protect the devices. Insets are given where applicable to show the potentiation (black) and depression (red) in more detail. In comparison, the undoped SiO$_2$ shows no stable potentiation with the chosen series resistor. The depression is dependent on delay times during potentiation. Short delays form more stable filaments, whereas long delays form "softer" filaments, easier

to depress. The potentiation and depression kinetics are mirroring the SET kinetics observed for same memory cells as shown in Figure 4. The experiment demonstrates how device performance and functionalities can be adjusted materials' design.

To demonstrate the influence of different dopants on potentiation and depression, we compared the responses of various devices applying pulse trains of 1 ms length with an amplitude of 1 V for the potentiation and 250 mV for the depression. The delay time was varied between 10 µs – 100 ms. Figure 6 shows the conductance of the undoped $SiO_2$, the (Al,Cu)-doped and the (Ga,Cu)-doped samples, respectively, where different delay times are sorted from 10 µs (lowest picture-units) over 1 ms (mid) up to 100 ms. For undoped devices, no controlled potentiation is possible, as the device switches already during the first pulse. The formed filament is (depending on the selected current compliance/resistance) unstable and dissolves within a short time period. This fast filament formation can be used to shorten the required pulse length and repetitions or to lower the magnitude of the learning voltage in order to decrease the programming power. On the other hand, the fast dissolution (at higher series resistance) is not favouring LTP, but makes these devices more suitable for fast spiking neural networks or threshold applications where the highly nonlinear nucleation regime is useful as filter for small pulse amplitudes. The internal driving force that enhances the fast self-driven RESET promotes this effect further.

Doped samples respond slower on potentiation (given by the larger capacitance and slower kinetics in low voltage regime) providing more flexibility for control on neuromorphic applications. The switching kinetics for intermediate voltages (0.5-2.5 V) is almost identical, leading to a much higher immunity to small voltage differences in the input pulses. This also improves significantly the ability to control the potentiation (speed and STP/LTP transition) by the delay time over several orders of magnitude. Additionally, the doping itself is enhancing the LTP function by stabilization of the metallic filament. The doping ensures initial concentation of Cu-ions, lowering their gradient within the $SiO_2$ matrix, As it can be seen in Figure 6 the applied depression pulses of 10 µs delays are not sufficient to decrease the conductance for doped devices. At 1 ms delay time, depression is observed for $SiO_2:(Al,Cu)_1$ device, while $SiO_2:(Ga,Cu)$ device shows still constant conduction. For 100 ms, full depression is reached fast for $SiO_2:(Al,Cu)_1$, while the Ga-doped sample is showing only initial stages of depression. Thus, doping is a very powerful tool to tune the responses to voltage signals in neuromorphic type learning functionalities. The learning characteristics can be additionally adjusted by variation of the pulse scheme, especially the voltage amplitudes in order to find optimum LTP and STP behavior for the targeted applications.

## Conclusions

In this work we demonstrate that materials' purity, respectively the chemistry and concentration of doping elements are essential factors in determining memristive functionalities in respect to all types of applications. Using $Cu/SiO_2/Pt$ as an example we observed that dopant concentrations as low as ppma range significantly change the device capacitance and switching/neuromorphic performance. By controlling the purity of the solid electrolyte and selecting the doping elements one can adjust the permittivity, electrochemical properties at the interfaces and in the electrolyte, and related device characteristics. Dopants determine the structure and overlapping of electrochemical double layers and the extent of internal charge separation. They slow down the switching kinetics due to interactions with mobile ions but increases the stability of the ON and intermediate states. These factors were identified as key elements for electric field accelerated reaction kinetics and transport at the nanoscale.

Doping is also a powerful tool to tune the ability for short- and long-term potentiation by improving the retention of the conductive filament and the (electrochemical) capacity.

Thus, we identify for first time the nature of the relation between materials properties and device functionalities and offer powerful design directions to engineer memristive materials for the whole spectrum of memristive applications.

## Experimental

Target synthesis

Our ultra-high purity silica films were specially prepared for the project by Heraeus GmbH, using synthetic fused silica with highest purity and no residual OH$^-$ content, so-called water-free silica[27].

To obtain thin electrolyte films with defined dopant concentration and lowest residual impurity concentration we developed new fused silica materials in the first step. Therefore, porous silica powder with high internal surface area was impregnated with different metal salt precursors of Cu, Al and Ga of nominal doping concentrations in an aqueous solution. These slurries are subsequently dried to obtain doped powders using a rotary evaporator also made of fused silica to achieve maximum doping homogeneity with lowest possible background impurity concentration. Doped powder samples are finally melted in a high-purity vacuum oven process to obtain the doped glasses which are finally cut and bonded into the final shape of the 2" sputtering targets [EP3208856A1], [EP2414872B1].

The obtained and tested targets with their corresponding dopant combinations are listed in Table 1.

Table 1: Sputter targets and its dopant concentrations.

| Sputtertarget | Dopant concentration(s) [ppma] | | remarks |
|---|---|---|---|
| SiO$_2$ | -/- | | 99,9999995 % SiO$_2$ |
| SiO$_2$:Cu | 2080 Cu | | |
| SiO$_2$:(Al,Cu)$_1$ | 14696 Al | 652 Cu | |
| SiO$_2$:(Al,Cu)$_2$ | 1443 Al | 45 Cu | |
| SiO$_2$:(Al,Cu)$_3$ | 142 Al | 5 Cu | |
| SiO$_2$:(Ga,Cu) | 8268 Ga | 39 Cu | |

The RF-magnetron sputtering was conducted in an off-axis configuration with rotating substrate to obtain homogeneous layers. Furthermore, the process gas mixture (Ar/O$_2$) and the substrate temperature were controlled to obtain dense SiO$_2$. For the rate determination and density of the sputtered films, X-ray reflectometry (XRR) was used, whereas SEM was used to reveal the microstructure of the films (Figure S1) The optimum parameter set was found to be: ≥ 150 °C substrate temperature, 2 µbar process pressure with a gas flow of 9 sccm Ar and 1 sccm O$_2$.

The tested samples were prepared in two different geometric configurations. For the capacitance measurements, 50 nm of the (doped) SiO$_2$ were sputtered with optimum parameter sets on a Si/SiO$_2$/Ti/Pt substrate as planar bottom electrode (30 nm Pt). A photolithography step followed by electron-beam evaporation of 30 nm Cu and DC-sputtering of 30 nm Pt to prevent excessive oxidation and finalizing lift-off step were performed to structurize top electrodes between 25 x 25 µm² and 1 x 1 mm². The equivalent schema of the setup is given in Supplementary Note 9.

For the kinetics and potentiation/depression measurements we structured a microcrossbar configuration with 5 x 5 µm² junction size on Si/SiO$_2$/TiO$_2$ substrates. Photolithography-assisted structurized platinum electrodes were DC-sputtered and subsequently covered with 10 nm of the to-be-tested SiO$_2$/dopant compositions via RF-sputtering. The bottom electrodes were kept partially open by another lithography step. Finally, a last lithography-assisted e-beam evaporation of 30 nm Cu and subsequent DC-sputtered Pt top electrodes completed the sample preparation.

The permittivities $\varepsilon_r$ were determined from capacitance measurements with a HP 4284A LCR meter. The voltages were set to 0 VDC and 20 mVAC with a frequency of 100 kHz. The permittivities were calculated according to $C = \varepsilon_0 \varepsilon_r \frac{A}{d}$, with the capacity $C$, the vacuum permittivity $\varepsilon_0$, the electrode area $A$ and the thickness of the oxide $d$. The $\varepsilon_r$-values were determined in air ($\approx$ 35 % relative humidity (RH)), vacuum ($p <$ 1e-4 mbar corresponding to 0 % RH) and wet nitrogen (> 90 % RH). The concentration of OH$^-$/H$_2$O in the film was calculated by the empirical formula $\varepsilon_s = 3.8073 + 2.72 * 10^{-22} \frac{cm^3}{ion} * N$ from [25], where $N$ is the concentration of hydroxyl groups per cm$^3$.

The SET kinetics were measured using a Wavetec 365 pulse generator and a Tectronix DPO 7254C Oscilloscope as depicted in Figure S7. The first channel of the oscilloscope was set to 50 Ω input impedance to record the pulse applied to the cell whereas the second channel was set to either 50Ω or 1 MΩ input impedance as shunt resistor. The voltage drop over channel 2 corresponds to the current flow through the cell.

The potentiation measurement in Figure 5 was performed with the same setup and 1 MΩ shunt resistance to observe RC-time implications clearly.

The potentiation and depression measurements in Figure 6 were performed with a Keithley 4200 using two pulse measurement units (PMU - each 50 Ω internal resistance) and a 100 kΩ resistance as current compliance to protect the devices. The pulses were applied via PMU-1 to the top Cu-electrode while the second PMU recorded the current response.

**Data availability**

The data that support the findings of this study are available from the authors on reasonable request, see author contributions for specific data sets.

**Acknowledgements**

The partial financial support by SF) 917 is gratefully acknowledged. The authors thank prof. R Dittmann for proof reading the manuscript and Thomas Poessinger for helping with graphical design.

**Author contributions**

C.N and I.V. contributed the idea and designed the study. M.L. designed and supervised the experiments. M.L., F.C., J.M. and M.v.W. performed the measurements and evaluated the data. R.W., C.N. and I.V. supervised the research. M.L. and I.V. wrote the manuscript. All authors contributed to the discussions of the results and improved the text.

Supplementary Information

**Design of materials properties and device performance in memristive systems**

**M. Lübben[1,2], F. Cüppers[1,2], J. Mohr[1,2], M. von Witzleben[1,2], U. Breuer[3], R. Waser[1,2,4], C. Neumann[5], I. Valov[2,4]**

[1]**Institut für Werkstoffe der Elektrotechnik II, RWTH Aachen University, Sommerfeldstraße 24, 52074 Aachen, Germany**

[2]**JARA – Fundamentals for Future Information Technology, 52425 Jülich, Germany**

[3]**Central Institute for Engineering, Electronics and Analytics, Forschungszentrum Jülich GmbH, 52425, Jülich, Germany**

[4]**Peter-Grünberg-Institut (PGI 7), Forschungszentrum Jülich, Wilhelm-Johnen-Straße, 52425 Jülich, Germany**

[5]**Heraeus Deutschland GmbH & Co. KG**

**Supplementary Note 1. Thin film deposition and characterization.**

In Supplementary Figure S1 the principle influences of the sputter parameters (substrate temperature and gas composition) in microstructure can be seen. The sputtering power was set to 150 W. The substrate temperature and the processing gas composition have a huge influence on the microstructure of the sputtered films. While the $SiO_2$ films are relatively porous if prepared at low temperatures, high substrate temperature deposition leads to a densification and improved homogeneity. An even larger effect is observed after addition of 10 % oxygen to the processing gas as can be seen comparing Supplementary Figure a) and d). The oxygen content leads not only to a tenfold reduction of the sputter rate (~15 nm/min down to 1,75 nm/min), but also for a significantly improved homogeneity and reduced roughness of the films. As a consequence we determined the optimum parameters for $SiO_2$ deposition to be at ≥150°C and a processing gas mixture of 90% Ar and 10 % oxygen.

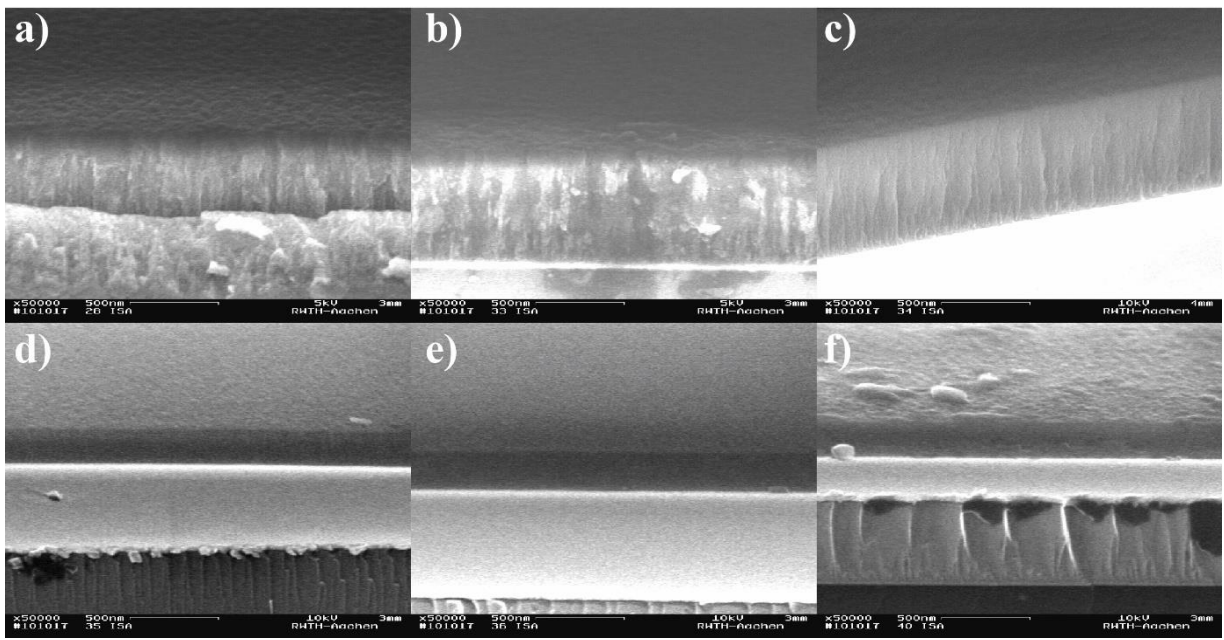

Supplementary Figure S1: Scanning electron microscopy images of sputtered $SiO_2$ films on platinum and its broken edges at 50.000 magnification. Deposition at a) Oxygen-free conditions at room temperature. b) Oxygen free conditions at 100 °C substrate temperature. c) Oxygen free conditions at 200 °C. d) 10% oxygen gas mixture at room temperature e) 10% oxygen gas mixture at 100 °C f) 10% oxygen gas mixture and 200 °C.

The dense and pore-free deposition process of stoichiometric $SiO_2$ allows to avoid/suppress external influences, such as the uptake and release of oxygen and/or moisture. The increase of the film density was also veryfied by XRR measurements (Supplementary Figure S2). For example, for an electron beam evaporated $SiO_2$ film of 30 nm thickness we extracted from the XRR a density of 1.95-2.0 g/cm³ whereas for RF sputtered films used in this study we obtained densities of ~2.2-2.25 g/cm³ regardless of the amount of dopant in the target.

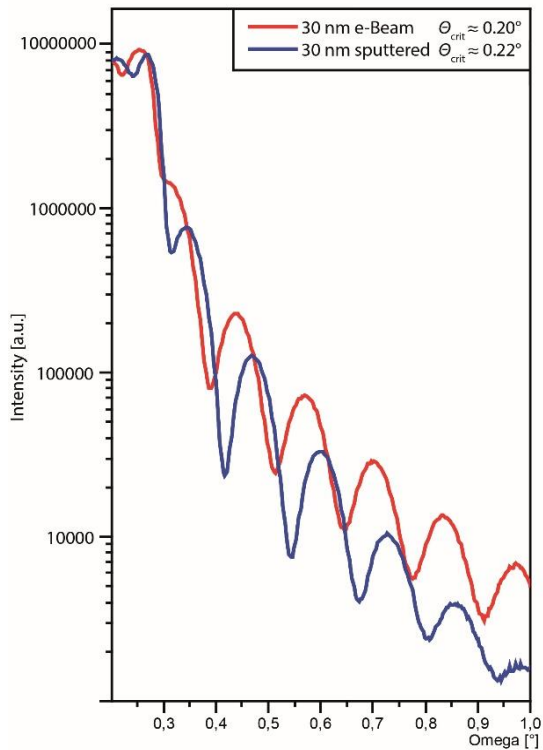

Supplementary Figure S2: Comparison of XRR measurements of electron beam evaporated (red) and sputtered $SiO_2$ films. The thicknesses of both films were determined to be ~30 nm, but have different critical angles $\Theta_{crit}$, therefore having different densities. The densities are determined to be $\rho_{e\text{-beam}}$ = 1.95-2.0 g/cm³ and $\rho_{sputtered}$ = 2.2-2.25 g/cm³, respectively.

The homogeneity of the compositions of the sputtered films were verified using secondary ion mass spectrometry (SIMS) depth profiling. An exemplary SIMS measurement of a highly doped $SiO_2:(Al,Cu)_1$ film is shown in Supplementary Figure S3. The intensities of all elements are constant throughout the whole film thickness.

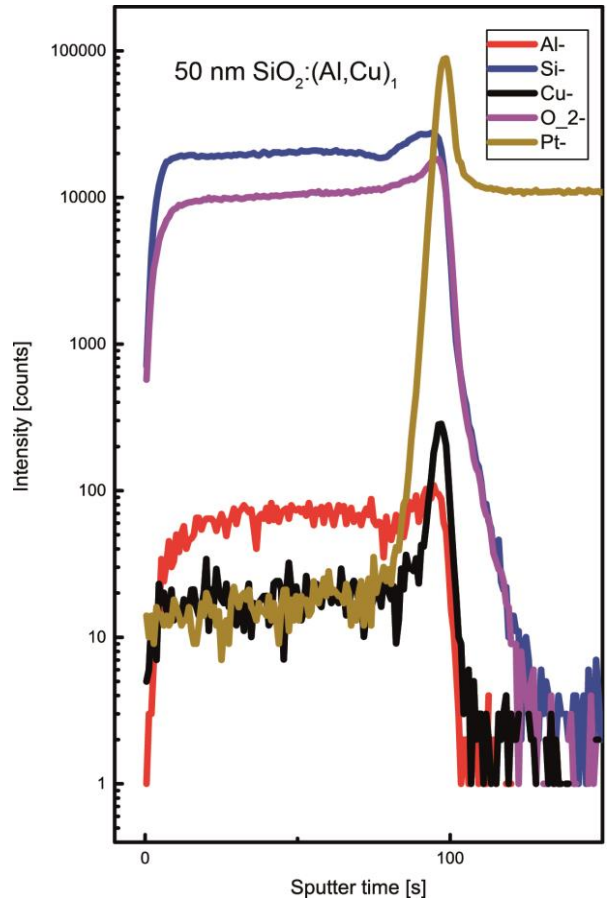

Supplementary Figure S3: SIMS depth profile of 50 nm sputtered $SiO_2$:(Al,Cu)$_1$ on platinized Si wafer substrate. The intensity of aluminum and copper mass signals is constant over the entire film thickness, verifying a homogeneous doping of the deposited film.

**Supplementary Note 2. Effects of moisture and materials' density.**

The sensitivity of determination of dopant concentration by permittivity measurements is high. As it can be seen in the inset of Supplementary Figure S4 even concentrations of 1 ppm to 20 ppm range can be effectively detected by this method.

In $SiO_2$ one can consider two different type of extrinsic defects – volatile and non-volatile. Protons and in general moisture as well other gases are considered volatile because in vacuum, in dry atmospheres and/or upon annealing these components leave the matrix. In contrary, Ga, Al and Cu remain in $SiO_2$. Of course, both type of defects can co-exists and each component present in the matrix will contribute to the change of the dielectric properties, irrespective whether these defects can potentially escape or not. Moreover, different dopants can also interact with each other.

Supplementary Figure S4 shows the change in dielectric permittivity as a function of non-volatile dopant concertation at conditions where the uptake of moisture in $SiO_2$ is close to zero (e.g. in vacuum).

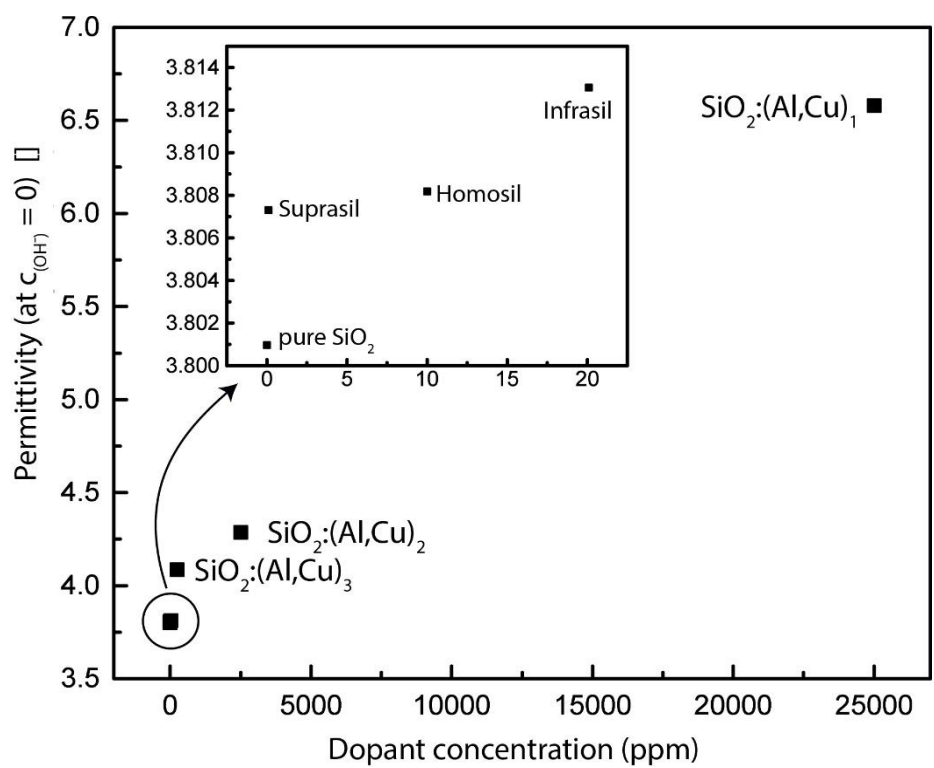

Supplementary Figure S4. Permittivity as a function of dopant concentration for moisture free samples. The inset is showing a zoomed region with the permittivity for industrial $SiO_2$ qualities Suprasil, Homosil and Infrasil with very low dopant concentrations, and ultra-pure $SiO_2$.

The permittivity increases linearly with increasing the non-volatile dopant concentration. Incorporation of moisture and/or $OH^-$-ions increases drastically the permittivity as water itself has a large $\varepsilon_r$ value of 78-80 at room temperature (OH- and immobilized water can have a permittivity between 4 to 80, corresponding to the degree of immobility[1]), whereas defect free $SiO_2$ (e.g. Suprasil-W from Heraeus GmbH) has a $\varepsilon_r$ value of 3.8073 at 300 K [2,3]. Because of the different electron affinities of the foreign dopants and/or defects, stronger or weaker

(compared to silicon and oxygen) interatomic electrostatic interactions are induced. These interactions lead to formation of dipoles, influencing the permittivity and capacitance of the electrolyte film.

The $\varepsilon$-values were determined at 100 kHz in air ($\approx$ 35 % relative humidity (RH)), vacuum ($p <$ 1e-4 mbar corresponding to 0 % RH) and wet nitrogen ($>$ 90 % RH). The concentration of OH$^-$/H$_2$O in the film was calculated by the empirical formula[2]:

$$\varepsilon_s = 3.8073 + 2.72 * 10^{-22} \frac{cm^3}{ion} * N \tag{S1}$$

where $N$ is the concentration of hydroxyl groups per cm$^3$. Thus, we have measured for pure SiO$_2$ under vacuum conditions a permittivity of $\varepsilon = 4.03$, that is corresponding to 8.2x10$^{20}$ cm$^{-3}$ residual hydroxyl groups.

The moisture dependence of the film permittivity of undoped, (Al,Cu) doped and (Ga,Cu) doped samples is shown in Supplementary Figure S5. The undoped and (Al,Cu) doped films have a similar increase in permittivity with higher OH$^-$ concentration levels, while the (Ga,Cu)-doped films have a flatter slope of the linear relation increase, indicating weaker electrostatic interactions between the incorporated water and the thin film.

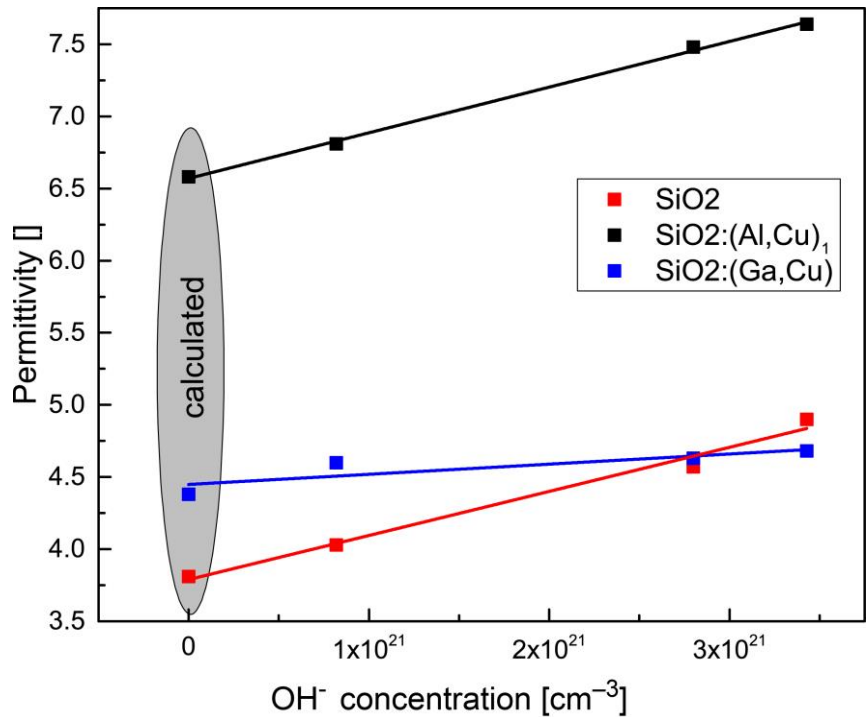

Supplementary Figure S5: Comparison of the permittivity dependence of hydroxyl concentration for pure $SiO_2$, highly doped $SiO_2$:$(Al,Cu)_1$ and doped $SiO_2$:$(Ga,Cu)$.

Moisture is not only changing dielectric properties but is also an essential factor for the electrode processes, as it can undergo the required counter electrode reactions at the counter electrode interface. This process is essential for the oxidation of the active electrode material in ECM[4, 5] or the incorporation of oxygen in VCM systems[6]. The counter redox reaction can also influence the switching performance as rate-limiting step[7].

We identified nanoscale porosity as one of the crucial factors, influencing the uptake of moisture. For example, RF-sputtered samples of pure $SiO_2$ used in this study have permittivities $\varepsilon_r$ ranging from 3.8 (dry - calculated) to 4.9 (wet) which corresponds to OH⁻ concentration of $3.4 \times 10^{21}$ cm⁻³ in the film. However, electron-beam evaporated samples (higher porosity, indicated by lower density) show a tremendous dependence of the permittivity varying between $\varepsilon_r \approx 5$ in vacuum and $\varepsilon_r > 12$ in moisture-saturated atmosphere.

**Supplementary Note 3. Space charge layer and potential/filed distribution in VCM cells**

For valence change memory devices (VCM) the general picture is not changing compared to ECM cells. More specifically, in Supplementary Figure S6 is presented the situation for only one mobile ionic specie i.e. oxygen vacancies (in terms of Kroeger-Vink notation). In this case the charge accumulation will occur only at one of the electrodes (in this case Ta). The other part of the oxide is depleted of oxygen vacancies and more insulting. Of course, considering moisture or $Ta^{5+}$ ions as additional ionic species, the profiles will take the form similar to that one shown in Figure 3.

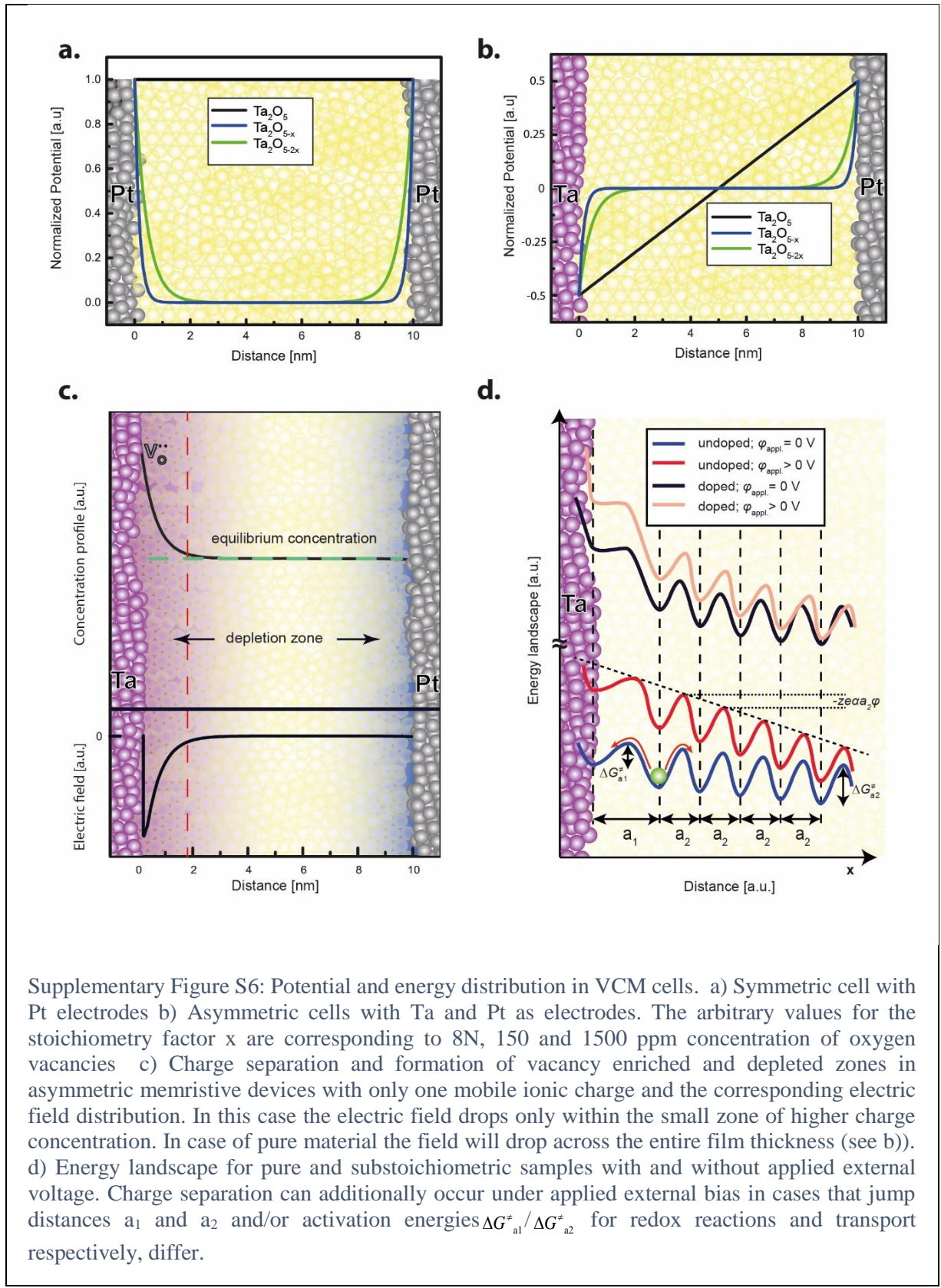

Supplementary Figure S6: Potential and energy distribution in VCM cells. a) Symmetric cell with Pt electrodes b) Asymmetric cells with Ta and Pt as electrodes. The arbitrary values for the stoichiometry factor x are corresponding to 8N, 150 and 1500 ppm concentration of oxygen vacancies c) Charge separation and formation of vacancy enriched and depleted zones in asymmetric memristive devices with only one mobile ionic charge and the corresponding electric field distribution. In this case the electric field drops only within the small zone of higher charge concentration. In case of pure material the field will drop across the entire film thickness (see b)). d) Energy landscape for pure and substoichiometric samples with and without applied external voltage. Charge separation can additionally occur under applied external bias in cases that jump distances $a_1$ and $a_2$ and/or activation energies $\Delta G^{\neq}_{a1}/\Delta G^{\neq}_{a2}$ for redox reactions and transport respectively, differ.

As seen from the field/potential profiles the existence of enriched and depleted zones will have significant impact on the switching kinetics in terms of both thermodynamic and kinetics. In highly pure samples the ions experience the filed effect across the entire film thickness, whereas in samples with high concentration of charges the field drops only within the enriched part, leading to a slower kinetics.

**Supplementary Note 4. Calculation of the Debye length.**

The Debye lengths $\lambda_D$ for the differently doped $SiO_2$ films are calculated corresponding to eq. S2

$$\lambda_D = \sqrt{\frac{\varepsilon_0 \varepsilon_r k_B T}{\sum_{i=1}^{N} N_A e^2 c_i z_i^2}}$$

(S2)

with the vacuum permittivity $\varepsilon_0$, the relative permittivity $\varepsilon_r$, the temperature $T = 293{,}15$ K, the Avogadro constant $N_A$, the elementary electric charge e, the ionic species concentrations $c_i$, the corresponding charge number $z_i$ and $N$ being the number of ionic species. For the calculations only the metallic ions are taken into account. The results are listed in Table S2.

Supplementary Table S2: Calculated Debye lengths.

| Sputter target | Debye length [nm] |
|---|---|
| $SiO_2$ | 69000 |
| $SiO_2$:Cu | 0.22 |
| $SiO_2$:(Al,Cu)$_1$ | 0.16 |
| $SiO_2$:(Al,Cu)$_2$ | 0.41 |
| $SiO_2$:(Al,Cu)$_3$ | 1.26 |
| $SiO_2$:(Ga,Cu) | 0.19 |

**Supplementary Note 5. Effects of the parameters $a$ and $\Delta G_a^{\neq}$ at applied external voltage.**

Equation 2 is formally representing field dependent redox reaction and transport, having own activation barrier and jump distance. The electrode redox reaction can be also regarded as a jump of an ion from the outset metal layer and first oxide layer. The jump distance and activation energy for redox reaction and transport in the general case is expected to deviate from each other, whereas these parameters should remain the same for the ion jumps related to transport within the film, for the case of pure material. Incorporation of doping elements within the solid electrolyte can add a second jump distance/activation energy, particularly in the case that these are rate limiting.

In case of symmetric cells e.g. Pt/Cu:SiO$_2$/Pt $\Delta\varphi_{int}$ is expected to be 0 and therefore at open circuit conditions more complex charge separation is expected. However, after external voltage $\Delta\varphi_{appl}$ is applied, mobile charge carriers of opposite sign (and different mobility) will be attracted to the oppositely biased electrodes and charge separation will be induced. Considering the whole cell, the electroneutrality will be kept, but locally there will be regions enriched or depleted, respectively of mobile charges, resulting in different local electric field distribution.

In addition, variations in *a* and $\Delta G_a^{\neq}$ can also lead to inhomogeneity in the electric field distribution. The jump distance *a* will or can be different within the same systems for the following cases: i) *a (redox)* and *a (transport)*. *a (redox)* is the jump distance for the redox reaction i.e. this will be the distance from the metal electrode to the first layer of the oxide. This length is usually expected to differ from the jump distance for the ion during the transport in the oxide matrix *a (transport)*. ii) In case of chemical or structural inhomogeneity of the oxide matrix there may be more than one jumping distance. For example, if we incorporate Al-ions that attract stronger the Cu-ions (compared to intrinsic defects). In such a case Cu slower transport/jump and longer jump distance *a (transport)* can be defined as from Al-to-Al ion. Structural defects such as dislocations, voids etc as well as nanocrystalline nuclei within amorphous matrix and/or agglomerations of different stoichiometry and/or composition will lead to same effect. In all these cases the activation energy is also expected not to be equal. Thus, upon application of external voltage charges can pile up at the defect locations and locally change the electric filed distribution.

**Supplementary Note 6. Influence of the protective resistance on the recorded SET time.**

Analyzing and discussing the SET kinetics we have to consider one important external factor, i.e. the input resistance used to protect the devices from overshooting and damages. Using higher resistances (more reliable in protecting the device) on samples with high permittivity may result in SET time determined entirely from the time of charging the device capacitance. For example, measuring with 1 MΩ input impedance devices of 5 µm x 5 µm with a film permittivity of $\varepsilon_r = 6$, will result in a RC time of 130 ns. The voltage $\varphi_{internal}$ effectively driving the redox reactions and the ion drift is given by eq. S3.

$$\varphi_{int} = \varphi_{cap} = \varphi_{appl}(1 - e^{-t/\tau}) \tag{S3}$$

with $\varphi_{cap}$ being the voltage over the capacity, $\varphi_{appl}$ the externally applied voltage, the time $t$ and the RC time $\tau$.

This effect was observed for highly doped $SiO_2:(Al,Cu)_1$ devices. As it can be seen in Figure 4 the SET time after exceeding voltage pulses of 2 V remained constant (blue region), limited be the RC time. The higher doping concentration and induced charge separation is leading to a formation of electrochemical capacitor (see Figure 3c). This additional capacitance (~ 15 µF/cm$^2$) increases the switching time in the saturation region to about 1 ms. To avoid this effect we used for these samples 50 Ω input impedance, lowering the estimated RC time to ~6.10$^{-12}$ s, allowing to probe the actual SET kinetics of the devices.

**Supplementary Note 7. Approaching ultra-fast switching**

To investigate the switching speed of the devices further below 10 ns timescale we also applied 10 ns pulses with a Picosecond PSPL2600C pulse generator and measured the current response with a Tektronix DPO73304D real time oscilloscope. The setup is similar to the one used in [8]. To provide prober impedance matching at the contact pads the device was integrated into a coplanar waveguide structure. The applied voltage was set to 2 V. As the device has a high initial resistance the voltage of the applied pulse doubles over the devices stack resulting in 4V at the device. In Supplementary Figure S7 the voltage at the device and the current response are shown. At the beginning of the pulse a small capacitive current can be observed which corresponds to the charging of the device. The SET event occurs approx. 1.4 ns after the maximum of the capacitive current.

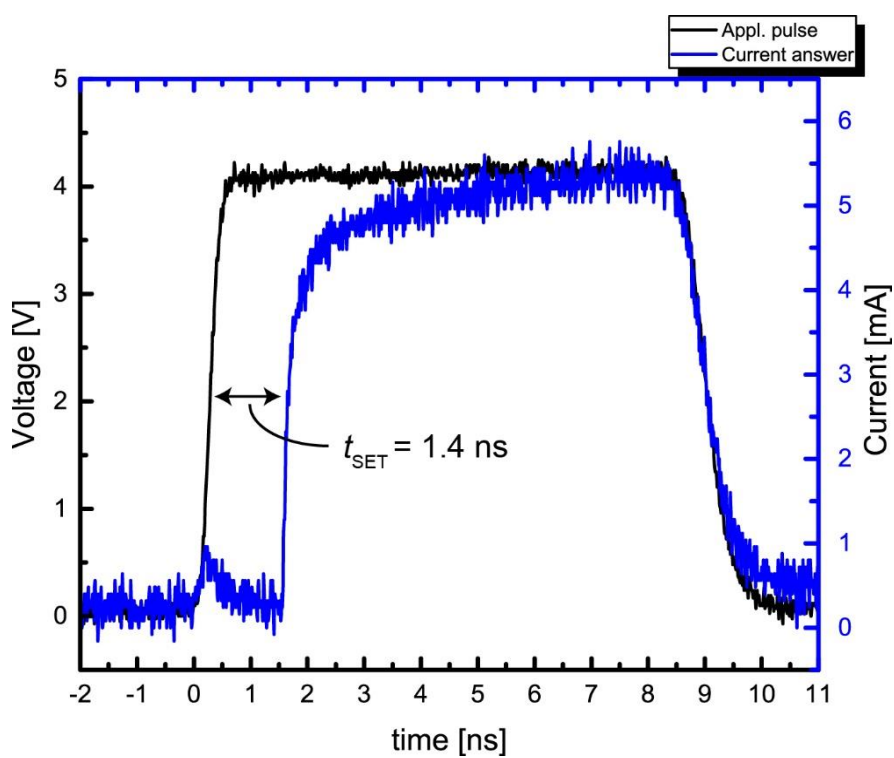

Supplementary Figure S7: Demonstration of ultra-fast switching of an undoped device with 3 nm thin SiO2. To investigate switching faster than 10 ns, this device was embedded in a coplanar waveguide structure and tested with a more sophisticated measurement apparatus including higher resolution and 50 Ω impedance matching.

**Supplementary Note 8. Dopant concentration dependent kinetics.**

The kinetics for all (Al,Cu) dopant concentrations (covering three orders of magnitude) are depicted in Supplementary Figure S8. In the low voltage regime ($V < 1$ V), the kinetics are almost congruent. For larger voltages the kinetics start to show different trends. The samples with the smallest dopant concentrations (SiO$_2$:(Al,Cu)$_3$) have a single slope up to the tested 5 V, independent on the input impedance. This indicates a weaker influence by the electrochemical capacitance and lower energy required to break the Al-Cu bonds. The medium doped SiO$_2$:(Al,Cu)$_2$ kinetics are also independent of the input impedance and the kinetics are similar to the highly doped SiO$_2$:(Al,Cu)$_1$ (with 1 MΩ impedance) up to 3 V. However, for higher SET voltages, the kinetics of SiO$_2$:(Al,Cu)$_2$ are becoming steeper for voltages > 3 V.

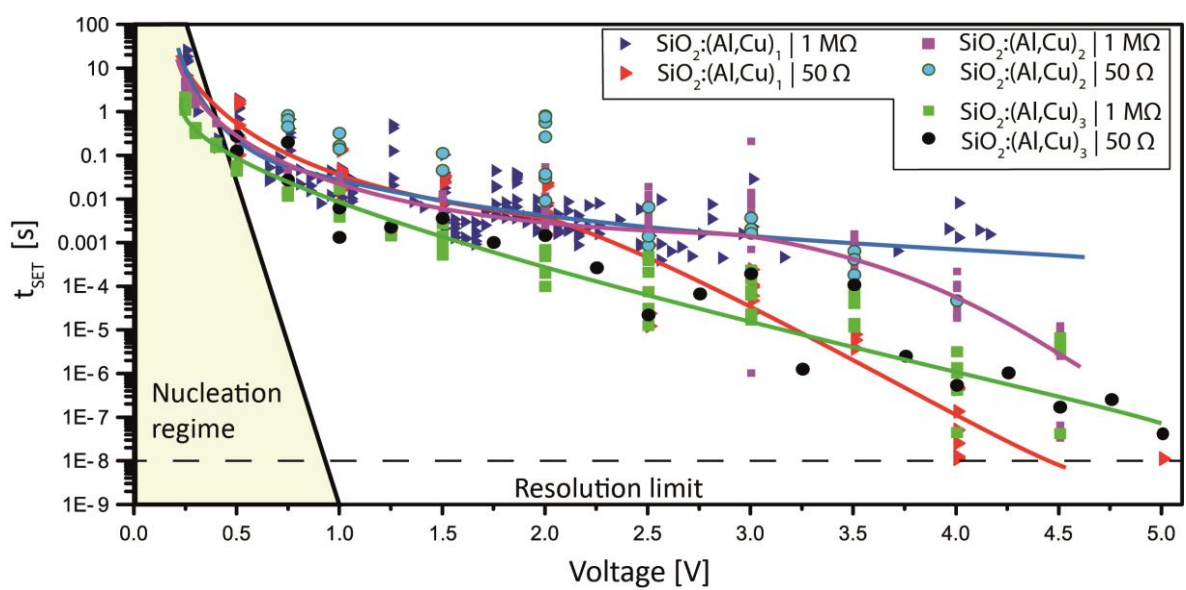

Supplementary Figure S8: Kinetics measurements with (Al,Cu)-doped samples and different concentration levels over 3 orders of magnitude.

**Supplementary Note 9. Measurement setup for switching kinetics.**

The choice for the input impedances of 50 Ω and 1 MΩ was made because of the easy setup and low influence of parasitic capacities (short cables and pulling the shielding of the cables on same potential right after the device). To protect the cells from irreversible damages, 1 MΩ input impedance of the oscilloscope was used. However, the high input impedance can hinder the investigation of the underlying physical processes (high RC time), and therefore we performed in parallel measurements with 50 Ω shunt resistance. Using 50 Ω input impedance the setup allows for a small RC time, and the internal capacitance of the devices is quickly saturated (for example for $\varepsilon_r = 6$ the time is ~ $6 \cdot 10^{-12}$ s) and the electrochemical processes and filament formation can further proceed. Whereas, when measuring with 1 MΩ input impedance, the RC time is drastically increased, the capacitance is loaded much slower (130 ns for same device), meaning that there will be a much longer time shift (voltage divider – equation S2) before switching processes can start.

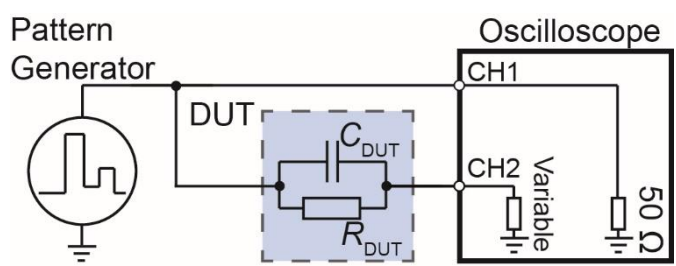

Supplementary Figure S9: Measurement setup for the SET kinetics determination. The pulse generator sends a rectangular pulse to the device simultaneously measured at a 50 Ω terminated channel (CH1) of the oscilloscope. The device response is measured at an either 50 Ω or 1 MΩ terminated second channel as converted voltage drop over the shunt resistor corresponding to the current flowing through the device. The SET time was determined to be the time difference between the rising edges at 50 % height.